# Finding the minimum energy conformation of protein-like heteropolymers by Greedy Neighborhood Search

**Joon Suk Huh**

## INTRODUCTION

The prediction of a protein's tertiary structure, given its amino acid sequence, is one of the intriguing problems in biophysics and computational biology. Anfinsen's thermodynamic hypothesis indicates that the tertiary structure of a protein is the minimum free-energy conformation [1]. Based on Anfinsen's hypothesis, the protein structure predicttion problem can be treated as a global optimization problem of its given energy function. Unfortunately, there is no analytical solution for this problem because of the complexity of the energy function.

Many global optimization methods such as Simulated Annealing (SA) [2], Genetic Algorithm (GA) [3], and Conformational Space Annealing (CSA) [4] have been proposed to deal with such hard optimization problems and successfully applied to various fields of science and engineering problems. A greedy algorithm is any algorithm that always takes the locally optimal solution while finding the global minimum of a function. Compared with above mentioned methods, it often fails to find the global minimum being trapped in a deep local minimum. A newly proposed global optimization method called Greedy Neighborhood Search (GNS) is belonged to the greedy algorithm class. GNS randomly generates the new solution adjacent to the current solution and then replaces the current one with the new one if the new one is better. Also GNS locates all the solutions on the minimum where the first derivative is zero and the second derivative is greater than zero by means of a local minimization algorithm. Because GNS randomly picks candidates on a minimum at each stage, it could escape from a deep local minimum unlike conventional greedy algorithms.

In this paper, the GNS method with a novel conformational sampling method using a spherical distribution called von Mises-Fisher distribution [12] is applied to find the minimum energy conformation of a protein-like heteropolymer model called AB model [5]. The AB model consists of only two types of monomers, Hydrophobic (A) and Hydrophilic (B) monomers analogous to the real proteins. Despite its simplicity, finding the minimum energy conformation of the AB model is very difficult and NP-complete problem. Unlike the traditional Hydrophobic (H), Polar (P) lattice model [6], only a few works were devoted to search the minimum energy conformations of the AB model polymers [7-11]. Because there is no rigorous proof to discriminate the true global minimum from many metastable local minima, the question is still open.

## MODELS AND METHODS

**The AB model**

The AB model polymers are modeled as an off-lattice bead-rod chain system in three-dimensional space shown in Fig. 1. The length of a rod connecting two consecutive beads is fixed to 1. The potential energy of an AB model polymer with N monomers is given by

$$E = \sum_{i=2}^{N-1} E_1(\theta_i) + \sum_{i=1}^{N-2}\sum_{j=i+2}^{N} E_2(r_{ij}, \zeta_i, \zeta_j) \quad (1)$$

where $E_1$ is the bond angle energy and $E_2$ is the non bonded van der Waals energy. The bond angle energy is given by

$$E_1(\theta_i) = (1 - \cos\theta_i)/4 \quad (2)$$

where $\theta_i$ is a bond angle defined by three consecutive monomers, $i-1$, $i$, and $i+1$. The van der Waals energy is given by

$$E_2(r_{ij}, \xi_i, \xi_j) = 4\left[ r_{ij}^{-12} - C(\zeta_i, \zeta_j) r_{ij}^{-6} \right] \quad (3)$$

where $r_{ij}$ is the euclidean distance between $i$ th monomer and $j$ th monomer. Each $\zeta_i$ is either A or B, and $C(\zeta_i,\zeta_j)$ is +1, +0.5 and -0.5 respectively, for AA, BB, AB pairs, producing strong attraction between AA pairs, week attraction between BB pairs, and week repulsion between AB pairs.

**A global optimization method**

The GNS method is performed on a search space consisting of local minima shown in Fig. 2. GNS consists of following five stages.

1. Random initialization – pick a random conformation from feasible space.

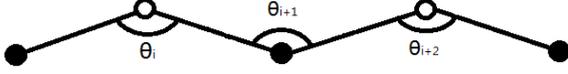

**Fig. 1** The bead-rod chain representation of the AB model. Each bead represents a hydrophobic (A) or a hydrophilic (B) monomer.

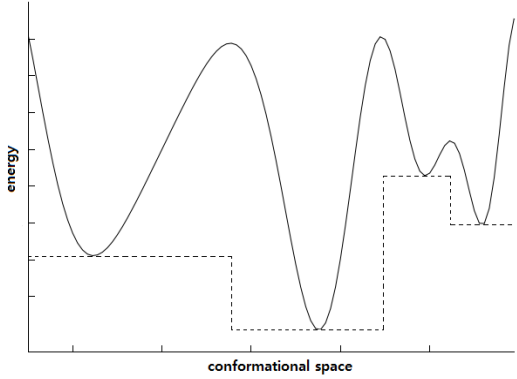

**Fig. 2** A search space consisting of local minima. All sampled conformations are relaxed by local minimization procedure.

2. Candidate generation – generate a trial conformation by a random modification on a current conformation.

3. Energy minimization – using a local minimization algorithm, relax the trial conformation.

4. Replacement – Replace the current conformation with the trial conformation if the trial conformation's energy is lower than the current one's.

5. Repeat from Stage 2 to Stage 4 until termination criterion (e.g. maximum iterations) is satisfied.

In this paper, L-BFGS subroutine [12] is used to relax the generated conformations.

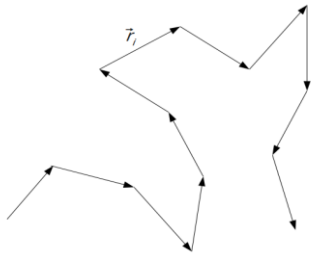

**Fig. 3** A chain system represented by edge vectors.

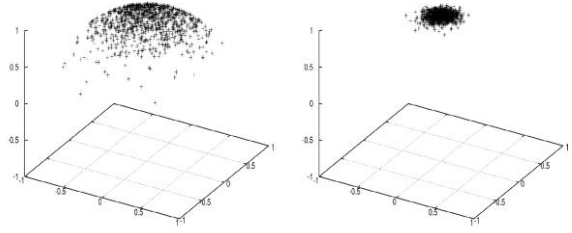

**Fig. 4** Points from the von Mises-Fisher distribution. The concentration parameters are 10 and 100 from left to right. The mean direction is $(0,0,1)^T$

**A conformational sampling method**

The bead-rod chain system with $N$ beads can be represented by $N-1$ edge vectors connecting two consecutive monomers as shown in Fig. 3.

The conformational sampling method used here generates a trial conformation by sampling these vectors from a spherical distribution called von Mises-Fisher distribution. The von Mises-Fisher distribution is a probability distribution of points or unit vectors on a $p$-dimensional hypersphere [13]. When $p$ is 2 (on the sphere), the probability density function is given by

$$p_3(x,\mu,\kappa) = c_3(\kappa) \cdot e^{\kappa \cdot \mu^T \cdot x}, \quad (4)$$

where

$$c_3(\kappa) = \frac{\kappa}{4\pi \sinh \kappa}, \quad (5)$$

where $\mu$ is the mean direction and $\kappa$ is the concentration factor. The greater the value of $\kappa$ is, the higher the concentration of the distribution around the mean direction is, as shown in Fig. 4.

In this paper, the method proposed by wood [14] is used to simulate the von Mises-Fisher distribution. A random unit vector $X$ sampled from the distribution with the mean direction $(0,0,1)^T$ and the concentration parameter $\kappa$ is given by

$$X = \sqrt{1-W^2} \cdot V + W \cdot \hat{z}, \quad (6)$$

where

$$W = \frac{1}{\kappa} \cdot \ln(e^{-\kappa} + \kappa \cdot C_3^{-1}(\kappa) \cdot y_1), \quad (7)$$
$$V = \cos\theta \cdot \hat{x} + \sin\theta \cdot \hat{y}, \quad (8)$$

where $y_1$ and $y_2$ are random variables from the uniform distribution over the range [0,1). A random unit vector with an arbitrary mean direction is obtained from $X$ by using following transfer matrices.

$$R_x = \begin{bmatrix} 1 & 0 & 0 \\ 0 & \cos\theta & \sin\theta \\ 0 & -\sin\theta & \cos\theta \end{bmatrix}, \quad (9)$$

$$R_z = \begin{bmatrix} \cos\varphi & \sin\varphi & 0 \\ -\sin\varphi & \cos\varphi & 0 \\ 0 & 0 & 1 \end{bmatrix}, \quad (10)$$

where $\theta$ is a polar angle of the mean vector and $\varphi$ is an azimuth angle measured from y-axis.

In this paper, trial conformation's edge vectors are generated from the von Mises-Fisher distribution whose mean directions are current conformation's edge vectors and the concentration parameter is 10.

## RESULTS AND DISCUSSIONS

The program is written in FORTRAN90 language. Experiments were conducted for four different sequences whose lengths are 13, 21, 34 and 55 respectively. The full sequences are given as

$$S_{13} = ABBABBABAB\ BAB,$$

$$S_{21} = BABABBAB \circ ABBABBABAB\ BAB,$$

$$S_{34} = ABBABBABAB\ BAB \circ BABABBA$$
$$\circ BABBABBABA\ BBAB,$$

and

$$S_{55} = BABABBAB \circ ABBABBABAB\ BAB$$
$$\circ ABBABBABAB\ BAB \circ BABABBAB$$
$$\circ ABBABBABAB\ BAB.$$

where the subscripts indicate the lengths of the sequences. For each sequence, ten independent simulations were performed on Intel C2D 2.0 GHz system with 2 GB memory using only one core.

Table 1 shows the minimum energies obtained by the GNS method, along with the results by nPERMis [7], ELP [8], CSA [10, 11] and STA [9]. For all sequences, results obtained by GNS are better than other methods for all the four sequences. It is quite surprising

|  | ME | MT | NS |
|---|---|---|---|
| $S_{13}$ | -5.1522 | 15.1s | 7/10 |
| $S_{21}$ | -12.9725 | 76.5s | 2/10 |
| $S_{34}$ | -25.5083 | 413.2s | 1/10 |
| $S_{55}$ | -42.7887 | 2520.8s | 1/10 |

**Table 1** mean energies (ME), mean CPU times in second (MT), and the number of times the minimum energy is found out of ten runs (NS).

that GNS can find much lower energy states than those of other methods despite its simplicity.

Fig. 5 shows the minimum energy conformations obtained by GNS, along with the high energy unfolded conformations, where orange spheres indicate hydrophobic monomers (A) and blue spheres indicate hydrophilic monomers (B). It can be seen that all conformations form single hydrophobic cores, which is analogous to the real protein conformations.

Table 2 shows mean energies and mean CPU times of ten individual runs along with the number of times the minimum energy is found. Even though the GNS method found the conformations whose energies are much lower than those of conformations obtained by other method, its success rate is lower than that of CSA. For CSA, the numbers of times the minimum energy found out of ten runs are 10, 10, 5, and 2 for $S_{13}$, $S_{21}$, $S_{34}$, and $S_{55}$ respectively. However, considering its relatively small number of iterations (100 times), success rate of GNS is acceptable.

## CONCLUSION

The lowest energy conformations of a protein-like AB model are found by Greedy Neighborhood Search. Also a novel conformational sampling method using von Mises-Fisher distribution is developed and applied to the candidate generation.

The energies obtained by GNS are much lower than those of other methods. This means that the results obtained by other methods are metastable local minima. Even though the AB model is not a real protein model, its geometrical representation is identical to the one-bead protein model unlike the HP lattice model. Therefore, if there is an adequate one-bead protein model that describes real protein's behavior, methods developed in this paper could be applied to that model directly.

|  | nPERMis [7] | ELP [8] | STA [9] | CSA1 [10] | CSA2 [11] | GNS |
|---|---|---|---|---|---|---|
| $S_{13}$ | -4.9616 | -4.967 | -4.9746 | -4.9746 | -4.9746 | -5.1888 |
| $S_{21}$ | -11.5238 | -12.316 | -12.3266 | -12.3266 | -12.3266 | -13.3492 |
| $S_{34}$ | -21.5678 | -25.476 | -25.5113 | -25.5113 | -25.5113 | -26.1121 |
| $S_{55}$ | -32.8843 | -42.428 | -42.5781 | -42.3418 | -44.7983 | -46.1154 |

**Table 2** The minimum energies obtained by Greedy Neighborhood Search (GNS) in comparison with those by nPERMis, ELP, STA, and CSA, respectively.

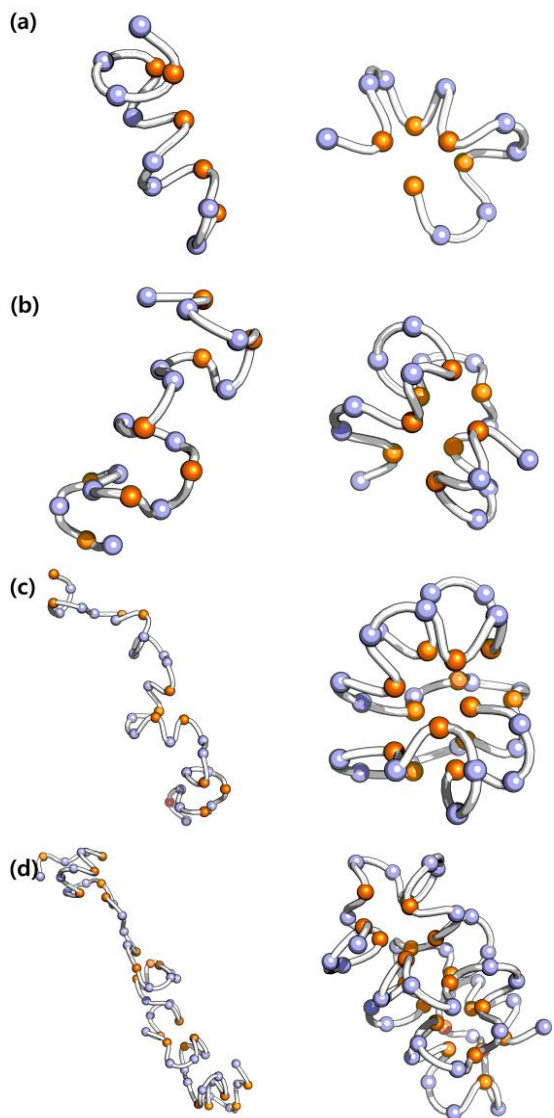

**Fig. 5** The minimum energy conformations (right), along with unfolded conformations (left) for (a) $S_{13}$, (b) $S_{21}$, (c) $S_{34}$, and (d) $S_{55}$.